# Perspectives of Ultra Cold Atoms Trapped in Magnetic Micro Potentials

József Fortágh, Sebastian Kraft, Andreas Günther, Christian Trück, Philipp Wicke, and Claus Zimmermann

Physikalisches Institut, Universität Tübingen, Germany

Abstract: Recent work on magnetic micro traps for ultra cold atoms is briefly reviewed. The basic principles of operation are described together with the loading methods and some of the realized trap geometries. Experiments that study the interaction between atoms and the surface of micro traps as well as the dynamics of ultra cold gases in wave guides are discussed. The results allow for an outlook towards future directions of research.

**Introduction**

After the spectacular success of the physics of ultra cold atoms, culminating in the Nobel Prize winning Bose-Einstein condensation, it is now a fascinating and rewarding challenge to find novel and promising applications for the newly available quantum gas. In this context, the recently introduced magnetic micro traps [1, 2, 3] offer particularly interesting perspectives. Consisting of simple combinations of current conductors that can be realized by standard micro fabrication techniques [4], they allow in principle, for constructing trapping potentials of almost arbitrary geometry. Conceivable are very elongated wave guide type potentials in which the atomic matter wave propagates in a single transverse mode similarly to photons in single mode optical fibers, but also interferometric arrangements based on spatial or temporal beam splitters are discussed [5, 6]. Such interferometers might possibly be combined to quantum gates feeding the vision of an "atom chip" [7] for quantum information processing. Before one can move towards such scenarios, the basic properties of magnetic micro traps have to be investigated in more detail in order to learn about their technical and fundamental limits. In this article, we summarize some of the recent results and try to contribute to the current discussion about micro traps by drawing preliminary conclusions which may help to give an outlook to the future of this very young research field.

**Basic Concepts of Magnetic Micro Traps**

We start with the basic idea of forming trapping potentials with inhomogeneous magnetic fields generated by thin conductors [1]. Magnetic trapping of atoms requires the atoms to be in a so called low field seeking state in which the atom's magnetic moment is oriented antiparallel to the local magnetic field at the position of the atom. Then, the mechanical action of an inhomogeneous magnetic field on a paramagnetic atom is given by a potential $U(\vec{r}) = \mu \cdot B(\vec{r})$ with the modulus of the magnetic moment of the atom $\mu$ and the modulus of the spatially dependent magnetic field $B(\vec{r})$. A simple trap can be constructed from an ordinary current carrying wire placed in a homogenous magnetic "bias field" $B_{bias}$ which is oriented perpendicular to the wire (Fig.1). On one side of the wire, its circular field is oriented anti-parallel to the bias field such that the two fields compensate along a line parallel to the wire. This line of vanishing magnetic field establishes the center of a magnetic wave guide where the atoms are radially confined but free to move along the axial direction. Near the center line of the wave guide, the leading term in the expansion of the magnetic field generates a potential U which increases linearly with the distance d from the center line: $U = \mu_B \cdot b \cdot d$. We refer to this wave guide as "quadrupole wave guide", characterized by the magnetic field gradient $b = \frac{\mu_0}{2\pi} I/B_{bias}$. For the magnetic moment of the atom $\mu$, we assume a value of one Bohr magneton $\mu_B$ which is realized by alkalies in the electronic substrate with the largest magnetic moment. For an infinitesimal thin wire, the gradient could be increased arbitrarily by simply reducing the current I in the wire. As a result the wave guide approaches the wire such that the thickness of a real wire eventually limits the gradient and with it the maximum value of the radial confinement.

Magnetic trapping only works in the adiabatic limit, i.e. if the local Larmor-frequency $\omega_L = \mu \cdot B/\hbar$ is larger than its relative temporal change as observed from the reference frame of the moving atom. For faster changes, the atom is lost due to Majorana spin flips. The atomic magnetic moment changes its orientation relative to the local magnetic field and the attractive trapping potential becomes repulsive [8]. At the center of a quadrupole wave guide the adiabaticity criterion is not fulfilled since the magnetic field vanishes and with it the Larmor-frequency. This problem can be fixed by adding a homogeneous "offset field" $B_0$ parallel to the wire (Fig.1). The total magnetic field modulus is then given by $|B| = ((b \cdot d)^2 + B_0^2)^{1/2} \cong B_0 + 1/2 \cdot (b \cdot d)^2/B_0$ and the atoms experience a harmonic radial potential: $U = \mu(b \cdot d)^2/2B_0 = 1/2 \cdot m \cdot \omega^2 \cdot d^2$. Here, m is the mass of the atom and $\omega = b(\mu/mB_0)^{1/2} = \mu \cdot b(m\hbar\omega_L)^{-1/2}$ the radial oscillation frequency in the "harmonic wave guide". A spatial variation of the offset field can now be used to additionally shape the potential along the wave guide. For instance, cross wires which add an inhomogeneity to the offset field can be introduced to terminate the guide at each end.

Large radial trap frequencies are desirable because they define the quantum mechanical level spacing for the modes of radial oscillation. A large level spacing facilitates the realization of quasi one dimensional propagation in the lowest transverse mode. We thus briefly discuss the maximum radial frequency that may be reached. In steep traps where the atoms move fast the Majorana spin flips are more pronounced and thus the maximum possible trap frequency depends on the desired life time of the atoms in the wave guide. For atoms in the vibrational ground state of a harmonic wave guide the Majorana loss rate can be calculated to be $\Gamma = \pi/2 \cdot \omega \cdot \exp(-\omega_L/\omega - 1/2)$ [8]. Since the trap frequency $\omega$ is connected to the Larmor-frequency (see above), the decay rate can be written as a function of the trap frequency alone $\Gamma = \pi/2 \cdot \omega \cdot \exp((\hat{\omega}/\omega)^3 - 1/2)$ with $\hat{\omega} = (\mu b/\hbar m)^{1/3}$ being a critical frequency that contains all the experimental parameters. For rubidium and a magnetic field gradient of b=100T/cm, which is close to the best experimentally achieved value [9, 10], $\hat{\omega}$ amounts to $2\pi \cdot 1.3 \cdot 10^6 s^{-1}$. In this case, the life time stays above 1 s for radial frequencies $\omega < 2\pi \cdot 100 kHz$. For higher frequencies the life time drops dramatically and reaches values of a few microseconds already for $\omega > 2\pi \cdot 200 kHz$. The frequencies in state of the art magnetic traps are thus comparable to the values obtained in optical lattices. This result depends only little on the atomic species since the mass enters the critical frequency only as a cubic root. As we will see later, surface effects may require to operate the wave guide at a larger distance from the conductor which reduces the achievable gradient. At then realistic values of b=1 T/cm, $\hat{\omega}$ reduces to $2\pi \cdot 6.2 \cdot 10^4 s^{-1}$ and the radial frequency must now be smaller than of $\omega = 2\pi \cdot 4.5 kHz$ to still allows for a lifetime of more than 1 s. In this case, the radial confinement is comparable to those obtained in the best optical traps generated by focused laser beams.

In practice, the current conductors are realized either by simple copper or aluminum wires [3, 11] or by micro fabricated gold, copper, or silver conductors. Micro fabrication, in principle, allows for extremely narrow conductors on the order of a micrometer and below, with a thickness of only several hundred nanometers and correspondingly large gradients of 100 T/cm and more [12, 9, 13, 14, 15]. However up to now, typical experiments were usually carried out with much wider conductors on the order of several 10 micrometers where micro fabrication is simply a method for making conductors with a well defined geometry. For calculating the trapping field of wide conductors the finite geometry has to be taken into account. Details are found elsewhere [10].

The basic arrangement of a single conductor in a bias field can be easily extended to more complex structures by using additional conductors. For example, in a combination of three parallel conductors, the outer pair may generate the bias field (Fig.2.a) and the wave guide is formed by a field that is generated at the chip only. Even simpler, a pair of parallel conductors driven with the same current already generates a useful wave guide without the need of an external bias field. However, its center line is positioned half way between the wires and, if micro fabricated conductors are used, a bias field is required to shift the center line from the surface of the substrate (Fig.2.b). This very simple geometry provides an interesting feature that may be exploited for the construction of a temporal interferometer [5]: If the bias field is continuously increased, a second line of minimal field modulus approaches the surface from infinity and merges with the original wave guide at a certain critical value of the bias field. Together they form a linear hexapole before they again split into two separated wave guides which move towards each of the two conductors [16] as illustrated in Figure 2.c. The de-Broglie wave function of an atom which is initially trapped at a small bias field in the minimum between the two conductors can be split into two parts by simply increasing the bias filed. After some holding time the bias field can be reduced again and the two parts of the wave function will recombine. However, if during the waiting time the relative phase between the two parts changes by a value of $\pi$ due to some extra potential, the two parts will recombine into the first radially excited state. Such an interferometer would realize a sensitive potential gradiometer, i.e. a force detector. Experimentally, the state selective single atom detection is a difficult task, however, not unconceivable. The realization of a spatial Mach-Zender type interferometer is another obvious geometry. In fact, the demonstration of a spatial beam splitter has been attempted by several groups. While for an incoherent cloud of thermal atoms a V-shaped intersection of conductors bears no particular difficulties [17], this approach is not successful for the coherent ensemble of a Bose-Einstein condensate [18]. As discussed later, a condensate turns out to be extremely sensitive to any small spatial variations of the wave guide potential. Thus, hitting the splitter results in violent and uncontrolled excitations of the condensate. For a smooth separation of a condensate a tunneling contact, analogous to optical fiber couplers, is probably more promising. Such a contact requires extremely fine structuring on a spatial scale of below one micro meter which is at the limit of what is possible today. On a larger scale, potential structures along the axis of the wave guide have already been used to control the position of the atoms inside the guide. They are, in practice, generated by the magnetic field of current conductors perpendicular to the wave guide. If a series of such conductors is combined to form a conveyor belt, the atoms can be moved along the surface of the chip (Fig.3). In a prototype experiment with a different geometry, a condensate has been shifted over a distance of 1.6 mm [19]. In this experiment, all conductors were placed at the front side of the chip, at the expense of a periodically changing curvature of the moving well in which the atoms travel. This may lead to unwanted excitations of the condensate which can be avoided at an optimized but more complex microstructure using a dual-layer technology illustrated in Figure 3. The operation of a dual layer conveyor belt with a cloud of ultracold thermal atoms is demonstrated in Figure 4. The atoms are transported over more than one centimeter and their position is well described by the simulation. The simulation also predicts an extremely smooth transportation with neglecting variation of the axial and radial trap frequencies [20]. The conveyor belt geometry can be extended in a natural way to generate a two dimensional periodic potential based on a grid of conductors. A recent experiment follows this approach and has shown the feasibility for thermal atoms and lattices with a relatively large lattice constant of 1.27 mm [21]. Reducing the dimensions of micro fabricated traps and approaching the scale of optical lattices is certainly a fascinating goal. As compared to optical lattices, the lattice constant is set by the geometry of the field generating elements offering unique possibilities for addressing single lattice sites and forming lattices with complex bases. Finally, one should not forget to mention a simple and popular way to accomplish axial confinement by using z-shaped conductors [7]. The bend conductors at each end of the wave guide generate an axial increase of the magnetic field modulus which terminates the guide potential and prevents the atoms from escaping. However, such a geometry is not very flexible in respect to variations of the axial and the radial oscillation frequency which compromises the initially attractive simplicity of the set up.

**BEC in the Micro Trap**

In first experiments, a canonical procedure for loading the micro trap with atoms was not obvious and several methods had to be developed. Today, most experiments use a so called surface magneto-optical trap (surface MOT) [4]. This very elegant and compact method requires the chip to act as a mirror that is placed in the center of a conventional MOT geometry. The optical set up of a MOT is compatible with the introduction of such a mirror and atoms can be collected and cooled close to the surface of the chip in very much the same way as in a conventional MOT. After the loading phase of the MOT the atoms are shifted closer to the chip surface by enhancing the magnetic quadrupole field of the MOT with conductors placed either behind or directly on the chip. Then, the MOT is turned off and the atoms are stored in the magnetic potential of the chip. In alternative setups, the ultracold cloud is actively transported from a region where a conventional magneto-optical trap is operated under optimized conditions into the micro potentials of the chip. These can be located at large distances from the MOT, even up to several cm. The transfer is accomplished either by shifting the center of the magnetic trap i.e. by a set of extra coils [22, 23], by optical tweezers [24] or by a magnetic wave guide [25]. These methods, although experimentally more sophisticated, offer the extra advantage of an up to two orders of magnitude larger number of atoms that can be loaded into the micro trap [23]. The separation of the chip from the optical region of the MOT also sets less constrains for the geometry and the surface of the atom chip (Fig.5).

After having learned how to load a micro trap, the tight confinement and the corresponding fast thermalization rate of the trapped gas facilitates further cooling into the regime of quantum degeneracy by forced evaporation, a standard method for reaching Bose-Einstein condensation [22, 10]. Today, Bose-Einstein condensates are routinely studied in magnetic micro traps by a number of research groups [22, 19, 18, 11, 26, 27, 15, 28, 29]. In the remainder of the article we discuss two aspects which are subject to current experiments. The first addresses the interaction between the micro traps surface and the atoms which become important at close distances, typically below 100 micro meter. Secondly, we describe some basic observations about condensates propagating in wave guides.

**Surface Effects**

One of the first experimental tests in exploring the properties of magnetic micro traps aims at the largest possible radial compression. It is not only limited by the intrinsic process of Majorana-losses which may occur while approaching the conductor and thereby increasing the oscillation frequency. Also material properties of the trap surface set significant limits. Loss and heating mechanisms have been predicted to be relevant for distances below a few micrometer [30]. Temporal fluctuations of the charge density inside the conductor at room temperature generate a time dependent magnetic field which induces atomic spin flips. The relative orientation between the magnetic moment and the local magnetic field is inverted and the potential well transforms into a potential hill. Consequently, the atoms are radially expelled from the wave guide. These losses have been studied in three recent experiments [11, 27, 15] with the result, that the atomic life time is reduced to below one second if the atoms approach the surface closer than approximately 4 micrometer. The loss being induced by Johnson-noise can significantly be reduced if the conductor is cooled down below the Debye temperature, i.e. to the temperature of liquid Helium. If a metallic conductor operated below the critical temperature for superconductivity, a gap in the excitation spectrum of electrons appears and thermal fluctuations are reduced exponentially. Using superconductors, even resistive heating could be avoided. Eventually, van-der-Waals and Casimir-Polder potentials will limit the nearest trap-surface distance to a few hundred nanometers [15]. Near the surface these attractive potentials increase with the sixths and fourth power, respectively, such that the quadratic power law of the magnetic wave guide potential is not strong enough to prevent the atoms from falling onto the surface.

With these predictions in mind it was quite a surprise when a fragmentation of atomic clouds was observed already at trap-surface distances on the order of hundred micrometers. Apparently, an unexpected axial potential splits an elongated atomic cloud into fragments with a typical separation on the order of 0,2 mm to 0,3 mm [31, 18] (Fig.6). We were able to show that the potential is of magnetic origin and that it depends on the sign of the current in the conductor [32]. In this experiment, a fragmented cloud is first prepared in an elongated wave guide and then the offset field is suddenly inverted. After letting the atoms adapt to the new field configuration for some milliseconds, their density distribution is recorded by absorption imaging. The positions of maximal density appears now at a position where the minima of the initial distribution have been observed, and vice versa. These findings can be explained by assuming a locally varying magnetic field component which is oriented parallel to the conductor. It is superimposed to the offset field and reduces or enhances the total local offset field depending on its local orientation. If the orientation of the offset field is flipped, the positions of enhancement and reduction are exchanged, transforming a minimum into a maximum and vice versa. The same behavior was observed where the current in the conductor has been inverted which allows for the conclusion that the unexpected field is generated by the current in a sign sensitive way. These results have been supported by experiments with atomic clouds brought close to the surface by optical tweezers [33]: fragmentation was detected only if a current was present in the conductor. Another experiment that also supports the magnetic origin of the fragmentation has measured the decay of the potential fluctuations with distance to the surface. It was found that it obeys a Bessel function which one would expect from a potential that originates from a meandering effective current inside the conductor [34]. The first convincing clue about the origin of the magnetic field fluctuations was given in a recent experiment [28]. In a careful study, the spatial fragmentation of the atomic cloud was mapped to the thickness fluctuations of the conductor and a clear relation between those has been found. Due to the thickness fluctuations, the mean current averaged over the cross section of the conductor deviates from a straight line which results in a radial current component responsible for the unwanted axial magnetic field. Calculations show that a deviation amplitude of the meandering current of only a few ten nanometers already leads to the observed potential structure [35]. The potential dimples that fragment a thermal cloud are sufficiently deep to trap a condensate [31]. Thus, it will be a technological challenge to construct long high quality wave guides with tight confinements. It requires an extremely precise control of the conductors geometry. On the other hand one may draw the maybe surprising conclusion, that ultra cold atoms make a very sensitive surface probe. To our knowledge no other method is able to measure such tiny magnetic field components on a 10000 times stronger back ground generated by the usual circular magnetic field of the conductor.

**Dynamics of condensates in the Micro Trap**

A second class of experiments addresses the question of how an ensemble of ultra cold atoms propagates in the magnetic wave guide at a chip. For single thermal atoms, one can identify a regime where its total energy is smaller than the spacing between radial excitations in the trap: $E_{total} < \hbar\omega$. In this case, the radial atomic motion is frozen and the dynamics is given by the solution of a one dimensional Schrödinger equation that contains the axial potential structure along the wave guide. In the experiments, not single atoms but atomic ensembles are loaded into the trap and the fast internal collision rates allow for defining a temperature. Then, the exclusive occupation of a single transverse mode is guaranteed only if the temperature is much smaller than the transverse level spacing $k_B T < \hbar\omega$. At values for the level spacing of about $2\pi \cdot 10 kHz$, the temperature should be significantly smaller than 500 nK which is easily achieved. However, with these parameters one reaches the degenerate regime already with about 10000 atoms even in wave guides with very weak axial confinement corresponding to an axial level spacing of only $2\pi \cdot 1 Hz$. This raises the question of how a condensate will behave inside a wave guide. The condition for quasi one dimensional, single mode dynamics now requires that the level spacing exceeds the chemical potential of the condensate. This sets a significant constraint on the maximum curvature of the axial potential along the guide. With 10000 atoms and a radial frequency of $2\pi \cdot 10 kHz$ only very flat and homogenous guides with axial frequencies of less than $2\pi \cdot 1 Hz$ can realize this regime. In terms of the above described surface potential which introduces much larger curvatures, this requirement seems difficult to meet. In fact, the quasi one dimensional regime has been reached in micro trap experiments only with a strongly reduced radial confinement of about $2\pi \cdot 1 kHz$ [23, 36]. With the reduced atomic density, the chemical potential now drops below the radial level for a condensate that has been allowed to expand axially in a shallow potential with a level spacing of only a few Hz. In this case the fragmentation effect is not a problem because the reduced radial confinement allows for a much larger guide-surface distance. This scenario identifies a regime of quasi one dimensional experiments which can be performed with micro traps, although the traps do not need to be very "micro". They can be constructed by simple wires, insolating grooves in conducting films [29], or similar techniques which are able to generate structures on the scale of about 0.1 mm.

The question of how such a quasi 1D condensates move in wave guides still remains. There are no comprehensive experimental data available yet, however we have performed a detailed study in the 3D regime which gives hints on the possible effects. With an atom number slightly exceeding the 1D regime, we have looked at the dynamics of a condensate in a wave guide which features a shallow but slightly anharmonic potential along the axial direction [37]. Surprisingly, the center of mass motion is almost undamped (quality factor > 20000) and can be precisely described with the model of a frictionless classical point mass in an anharmonic conservative potential (Fig.7.a). More interesting is the observed excitation of internal degrees of freedom. In the reference frame of the oscillating condensate, the confining axial potential periodically changes its curvature due to the axial anharmonicity of the wave guide. This results in a periodic squeezing of the condensate with the frequency of the center of mass motion. Consequently, the aspect ratio of the condensate (Fig.7.b) exhibits drastic temporal fluctuations with a nontrivial frequency spectrum (Fig.7.c). The observations can be perfectly simulated with the Gross-Pitaevskii equation and are thus well understood [38]. In our context here, it is important to note that the oscillations prevail even when, at the end of the oscillation period the number of atoms has dropped sufficiently to enter the quasi 1D regime. This means that single mode integrated atom optics in the degenerate regime have to deal with these excitations in one way or the other.

**Outlook**

For the future, one can proceed in different directions. One fruitful scenario compatible with the presently known limitations could be the use of a condensate mainly as source for non thermal atoms of well defined energy. In Figure 8, an experiment is shown with a condensate kept at one end of a wave guide by a suitable potential barrier. By slowly lowering the height of the barrier, a bunch of atoms is released into the guide where it rapidly expands [23]. During this expansion, the mean field interaction energy between the atoms is released resulting in an expanding wave packet of 10000 atoms all being in the same state which can approximately be described by a single atom Schrödinger equation. Such an "on chip atom laser" could be one part of an "atom spectrometer" which is completed by a single atom detector [40, 41] at the other end of the wave guide. In an optimized setup, more than 100000 atoms could be continuously sent through the guide within one second and assuming almost 100% detection efficiency of the detector, a similar number of data points could be acquired. As a simple sample for testing the performance of such a spectrometer, an additional box potential could be introduced into the guide between the atom laser source and the detector (Fig.9). For atomic energies just slightly above the height of the test barrier, the transmission is given by an Airy function and a matter wave analog to an optical Fabry-Perot spectrometer is realized. By varying the parameters of the test barrier, a complete spectrum can be taken with the atoms from a single condensate. Additional forces acting on the atoms will change the transmission and it will be interesting to see whether sensitive detectors can be constructed this way. Different kinds of probes could be investigated with the spectrometer as for instance atomic clouds of different species. Their influence can be described by an effective index of refraction which is determined by the elastic and the inelastic scattering properties. Besides looking at the number of the transmitted atoms, one can also detect the coherence properties by recording the second order coherence function of the atoms arriving at the detector. This might also be a viable approach for studying 1D-gases in the so called Tonks-Girardeau regime [42]. After having established the coherence properties of the laser output, their change induced by the sample could be investigated giving access to specific properties of the sample.

A second path for future investigation could be the use of a condensate as a surface probe. Already current experiments have demonstrated condensates in micro traps to act as very sensitive probe for static [32] and fluctuating magnetic fields [15]. The detection of induced electrostatic interaction is within reach [43]. Experiments with optical lattices are inspiring in a sense that they show how to visualize the reciprocal lattice by time of flight imaging techniques [44]. Since the size of the reciprocal lattice increases with decreasing lattice constant, a condensate that has been exposed to the surface of a sample could act as a microscope with a resolution which grows for ever finer spatial structures.

The experimental investigation of magnetic micro traps has just begun and an increasing number of research groups feel attracted to this young field at the interface between basic research and applied physics. By following the initial motivation of constructing extremely steep potentials, unexpected new perspectives have been discovered and there probably will be further surprises in the future. It will be fascinating to see where we will be in ten years from now.

**Acknowledgement**

We greatly acknowledge Herwig Ott for his contributions to experiments we present in this article as well as for the inspiring discussions and his successful research in our laboratory during several years. Our work presented in this review article has been supported in parts by the Deutsche Forschungsgemeinschaft and by the European Union.

**References**

[1] J.D. Weinstein, K.G. Libbrecht, Phys. Rev. A 52, 4004 (1995).
[2] V. Vuletic, T. Fischer, M. Praeger, T.W. Hänsch, C. Zimmermann, Phys. Rev. Lett. 80, 1634 (1998).
[3] J. Fortágh, A. Grossmann, T. W. Hänsch, C. Zimmermann, Phys. Rev. Lett. 81, 5310 (1998).
[4] J. Reichel, W. Hänsel, T.W. Hänsch, Phys. Rev. Lett. 83, 3398 (1999).
[5] W. Hänsel, J. Reichel, P. Hommelhoff, and T. W. Hänsch, Phys. Rev. A 64, 063607 (2001).
[6] E. Andersson, T. Calarco, R. Folman, M. Andersson, B. Hessmo, and J. Schmiedmayer, Phys. Rev. Lett. 88, 100401 (2002).
[7] R. Folman, P. Krüger, J. Schmiedmayer, C. Henkel, Adv. At. Mol. Opt. Phys. 48, 263 (2002).
[8] C. V. Sukumar and D. M. Brink, Phys. Rev.A 56, 2451 (1997).
[9] N. H. Dekker, C. S. Lee, V. Lorent, J. H. Thywissen, S. P. Smith, M. Drndic, R. M. Westervelt, and M. Prentiss, Phys. Rev. Lett. 84, 1124 (2000).
[10] J. Reichel, Appl. Phys. B 74, 469 (2002).
[11] M. P. A. Jones, C. J. Vale, D. Sahagun, B. V. Hall, and E. A. Hinds, Phys. Rev. Lett. 91, 080401 (2003).
[12] M. Drndic, K.S. Johnson, J.H. Thywissen, M. Prentiss, and R.M. Westervelt, Appl. Phys. Lett. 72, 2906 (1998).
[13] R. Folman, P. Krüger, D. Cassettari, B. Hessmo, T. Maier, and J. Schmiedmayer, Phys. Rev. Lett. 84, 4749-4752 (2000).
[14] J. Fortágh, H. Ott, G. Schlotterbeck and C. Zimmermann, B. Herzog and D. Wharam, Appl. Phys. Lett. 81, Nr. 6, 1146 (2002).
[15] Yu-ju Lin, Igor Teper, Cheng Chin, and Vladan Vuletic, Phys. Rev. Lett. 92, 050404 (2004).
[16] E. A. Hinds, C. J. Vale, and M. G. Boshier, Phys. Rev. Lett. 86, 1462 (2001).
[17] D. Cassettari, B. Hessmo, R. Folman, T. Maier, and J. Schmiedmayer, Phys. Rev. Lett. 85, 5483-5487 (2000).
[18] A.E. Leanhardt, A.P. Chikkatur, D. Kielpinski, Y. Shin, T.L. Gustavson, W. Ketterle, and D.E. Pritchard, Phys. Rev. Lett. 89, 040401 (2002).
[19] W. Hänsel, P. Hommelhoff, T.W. Hänsch, and J. Reichel, Nature 413, 498 (2001).
[20] A. Günther et.al., under preparation.

[21] A. Grabowski and T. Pfau, Eur. Phys. J. D 22, 347-354 (2003).
[22] H. Ott, J. Fortágh, G. Schlotterbeck, A. Grossmann, and C. Zimmermann, Phys. Rev. Lett. 87, 230401 (2001).
[23] J. Fortágh, H. Ott, S. Kraft, A. Günther, and C. Zimmermann, Appl. Phys. B 76, 157-163 (2003).
[24] T. L. Gustavson, A. P. Chikkatur, A. E. Leanhardt, A. Görlitz, S. Gupta, D. E. Pritchard, and W. Ketterle, Phys. Rev. Lett. 88, 020401 (2002).
[25] Dirk Müller, Eric A. Cornell, Marco Prevedelli, Peter D. D. Schwindt, Ying-Ju Wang, and Dana Z. Anderson, Phys. Rev. A 63, 041602(R) (2001).
[26] S. Schneider, A. Kasper, Ch. vom Hagen, M. Bartenstein, B. Engeser, T. Schumm, I. Bar-Joseph, R. Folman, L. Feenstra, and J. Schmiedmayer, Phys. Rev. A 67, 023612 (2003).
[27] D. M. Harber, J. M. McGuirk, J. M. Obrecht, E. A. Cornell, J. Low Temp. Phys. 133, 229 (2003).
[28] J. Esteve, Ch. Aussibal, Th. Schumm, C. Figl, D. Mailly, I. Bouchoule, Ch. Westbrook, A. Aspect, Phys. Rev. A 70, 043629 (2004).
[29] C. J. Vale, B. Upcroft, M. J. Davis, N. R. Heckenberg and H. Rubinsztein-Dunlop, cond-mat/0406015.
[30] C. Henkel, S. Pötting, and M. Wilkens, Appl. Phys. B 69, 379 (1999).
[31] J. Fortágh, H. Ott, S. Kraft, A. Günther and C. Zimmermann, Phys. Rev. A 66, 041604(R) (2002).
[32] S. Kraft, A. Günther, H. Ott, D. Wharam, C. Zimmermann, and J. Fortágh, J. Phys. B: At. Mol. Opt. Phys. 35, L 469 (2002).
[33] A. E. Leanhardt, Y. Shin, A. P. Chikkatur, D. Kielpinski, W. Ketterle, and D. E. Pritchard, Phys. Rev. Lett. 90, 100404 (2003).
[34] M. P. A. Jones, C. J. Vale, D. Sahagun, B. V. Hall, C. C. Eberlein, B. E. Sauer, K. Furusawa, D. Richardson and E. A. Hinds, J. Phys. B: At. Mol. Opt. Phys. 37 No 2. (2004)
[35] J. Fortágh, PhD Thesis, Universität Tübingen (2003).
[36] E. A. Cornell, private communication.
[37] H. Ott, J. Fortágh, S. Kraft, A. Günther, D. Komma, and C. Zimmermann, Physical Review Letters 91, Nr. 4 (2003) 040402-1.
[38] H. Ott, J. Fortágh and C. Zimmermann, J. Phys. B: At. Mol. Opt. Phys. 36 (2003) 2817-2822.
[39] Y. Castin and R. Dum, Phys. Rev. Lett. 77, 5315 (1996).
[40] P. Horak, B. G. Klappauf, A. Haase, R. Folman, and J. Schmiedmayer, P. Domokos, E. A. Hinds Phys. Rev. A 67, 043806 (2003).
[41] R. Long, T. Steinmetz, P. Hommelhoff, W. Hänsel, T. W. Hänsch, and J. Reichel, Phil. Trans. R. Soc. Lond. A 361, 1375 (2003).
[42] M. Girardeau, J. Math. Phys. (N.Y.) 1, 516 (1960).
[43] J. M. McGuirk, D. M. Harber, J. M. Obrecht, and E. A. Cornell, cond-mat/0403254.
[44] M. Greiner, I. Bloch, O. Mandel, T.W. Hänsch and T. Esslinger, Phys. Rev. Lett. 87, 160405 (2001).

**Figures**

Figure 1
Field configuration of a magnetic wave guide. A current carrying conductor generates a circular magnetic field which is compensated by a homogeneous bias field $B_{bias}$. The modulus of the resulting magnetic field has a minimum along a line parallel to the conductor which establishes a quadrupole wave guide for ultra cold paramagnetic atoms. With an additional offset field $B_0$ parallel to the conductor, the geometry of a harmonic wave guide is accomplished.

Figure 2
Qualitative sketch of magnetic field lines for wave guides at a chip. The orientation of the current is indicated by a cross and a dot, the centre of the wave guide is indicated by a filled circle. a) Wave guide generated by the field of three conductors. b) and c) Wave guides generated by two conductors and an external bias field. The position of the wave guides is determined by the relative strength of the conductors´ magnetic field and the bias field.

Figure 3
Dual-layer atom-chip. Three parallel conductors (C1, C2, C3) on the chip surface provide the magnetic field for the radial confinement of a wave guide. Perpendicular conductors at the back side of the 250 mm thin substrate produce the offset field and axial confinement. By changing the current in these cross conductors, the trap centre can be shifted along the wave guide. Between the conductors C1, C2 and C3, another set of conductors is placed allowing a variety of experimental sites with different micro trap potentials.

Figure 4
Forth and back transport of a thermal cloud 300 µm below the chip surface over a total length of 10.4 mm. a) Absorption images and b) position of the cloud centre (data points) during the transport. The solid line was calculated with the conductors geometry and the distribution of driving currents showing the expected position. Thus, a precise control of the trap´s position is demonstrated.

Figure 5
*In-vacuo* magneto-optical trap (MOT) and magnetic micro trap setup. The setup is placed in a vacuum chamber at a base pressure of $10^{-11}$ mbar. The MOT is generated at the right hand side in the magnetic quadrupole field of the MOT-coils and three pairs of counter propagating laser beams. The MOT is loaded from a thermal beam of rubidium atoms [23]. After collecting $10^8$ $^{87}$Rb atoms at a temperature of 100 µK, the laser beams are turned off and the atoms are restored in the magnetic quadrupole field generated by the MOT-coils. The trap is now shifted adiabatically to the left hand side to the middle of the transfer-coils by 34 mm. Because the two pairs of coils overlap, the transfer can be easily done by ramping up the field of the transfer-coils and ramping it down at the MOT-coils. The adiabatic transfer does not suffer from losses. On the left hand side, an Ioffe-trap is generated by superimposing the magnetic field of a vertical Ioffe-wire [3]. Here, the atomic cloud can be cooled by forced evaporation before shifting it up 2.2 mm to the micro trap at the microstructure. After an adiabatic compression, a few times $10^7$ atoms arrives in the micro trap where they are further cooled. Bose-Einstein condensation is reached with typically $10^6$ atoms at a critical temperature of 1 µK. In the micro trap, both the thermal cloud or the condensate can be manipulated. The spatial separation of MOT and micro trap combines two advantages: the efficient preparation of ultracold atoms in a three dimensional volume of a six beam MOT and the efficient cooling in the tight confinement of the micro trap.

Figure 6
Density distribution of a thermal cloud after expansion in a magnetic wave guide (absorption image and density profile). The density distribution of the cloud exhibits an axial potential modulation in the wave guide which is generated at a thin current conductor. The images a) and b) were taken for two opposite orientations of the offset field and show that inverting the offset field changes potential minima into maxima and vice versa. Thus the potential modulation is caused by an uncontrolled axial magnetic field component (see Text).

Figure 7
a) Centre of mass oscillation of the condensate after an initial displacement from the equilibrium position in an anharmonic wave guide (absorption images). b) Oscillation of the aspect ratio excited by the centre of mass oscillation (data points). The solid line shows the simulation with the Gross-Pitaevskii equation. c) Frequency spectra of the aspect ratio. The observed frequency components are in good agreement with the theory: (i) the fundamental trap frequency $\nu_0$, (ii) its second harmonics $2\nu_0$, (iii) the lowest lying eigenfrequency of the condensate $(5/2)^{1/2}\nu_0$ and mixed frequencies.

Figure 8
Releasing a condensate into a wave guide. The axial confinement of the magnetic trap was turned off within 400 ms and a gradient field forces the condensate into the wave guide. The release of the condensate into the wave guide is completed after 385 ms. a) Position of the condensate at different stages of the expansion. The absorption images were taken after 23 ms *time-of-flight* and show the propagation of the condensate as well as the change of the its shape indicating a reduction of the interaction energy. b) Time evolution of radial (open circles) and axial (filled squares) size of the condensate during the expansion. The solid line shows the numerical simulation of the expansion [39]. The out-coupling process is smooth enough that no collective excitations are observed.

Figure 9
A simple atom interferometer. Atoms are coupled out from a BEC reservoir into a box potential. They are transmitted to the single atom detector if the resonance condition of the interferometer is fulfilled. Thus, the box potential in the wave guide represents a sensitive momentum filter.

Figure 1

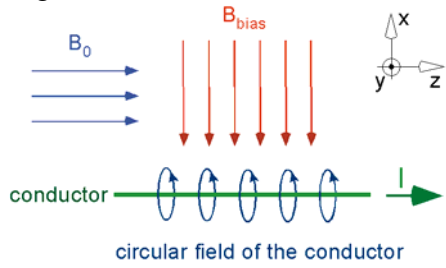

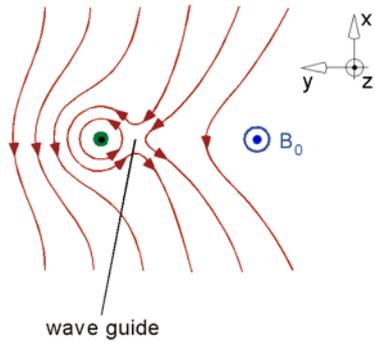

Figure 2

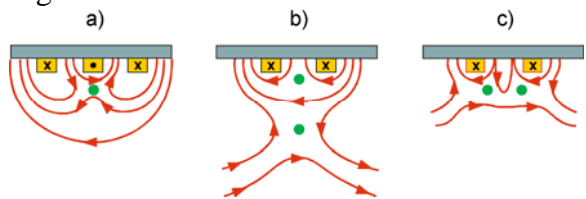

Figure 3

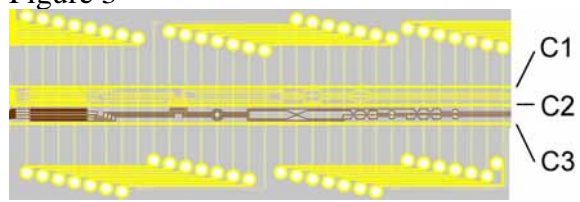

Figure 4

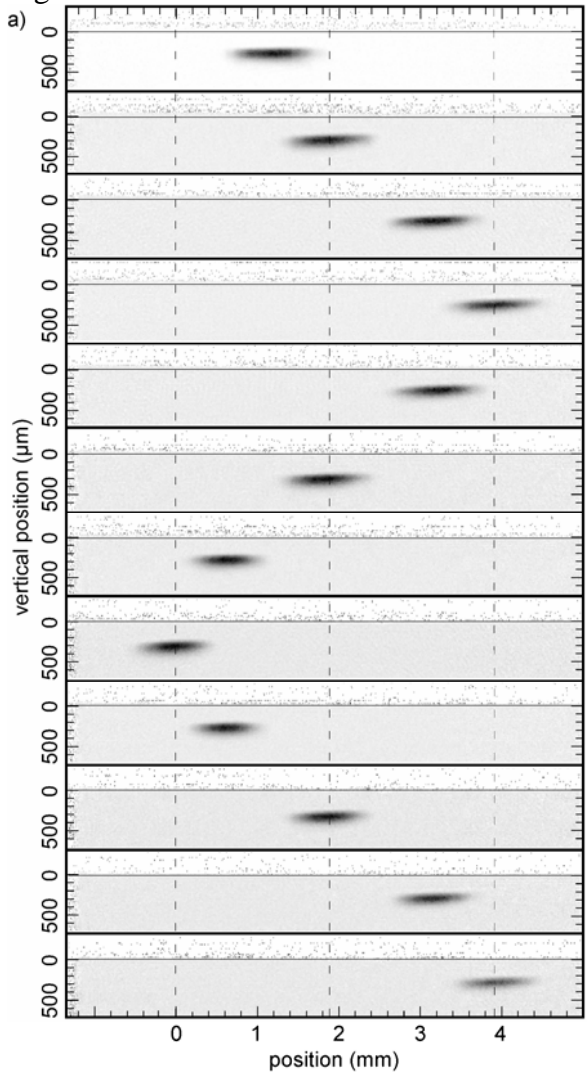

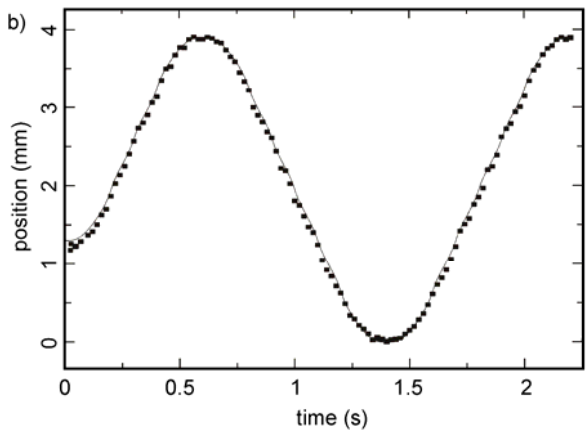

Figure 5

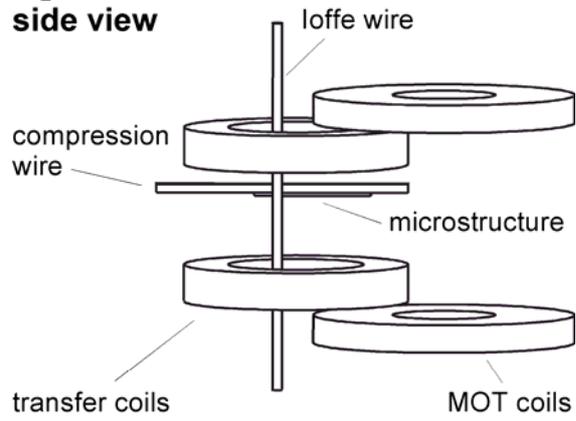

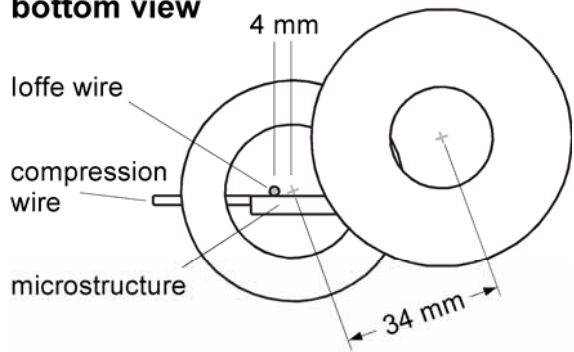

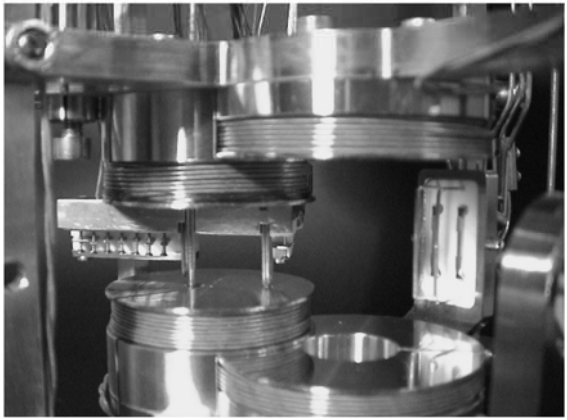

Figure 6

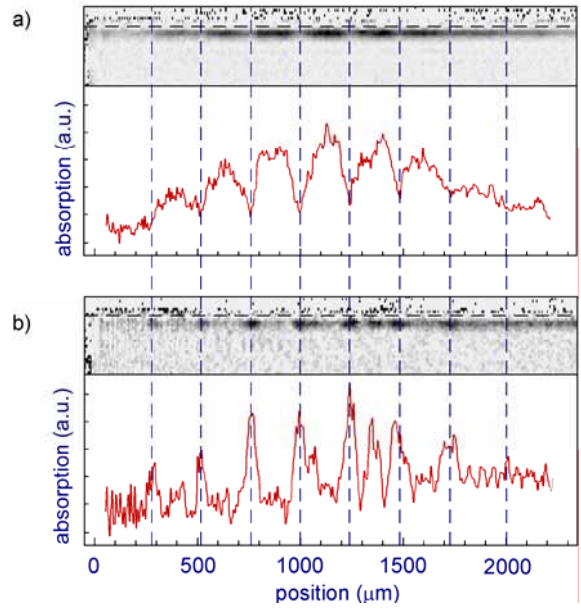

Figure 7

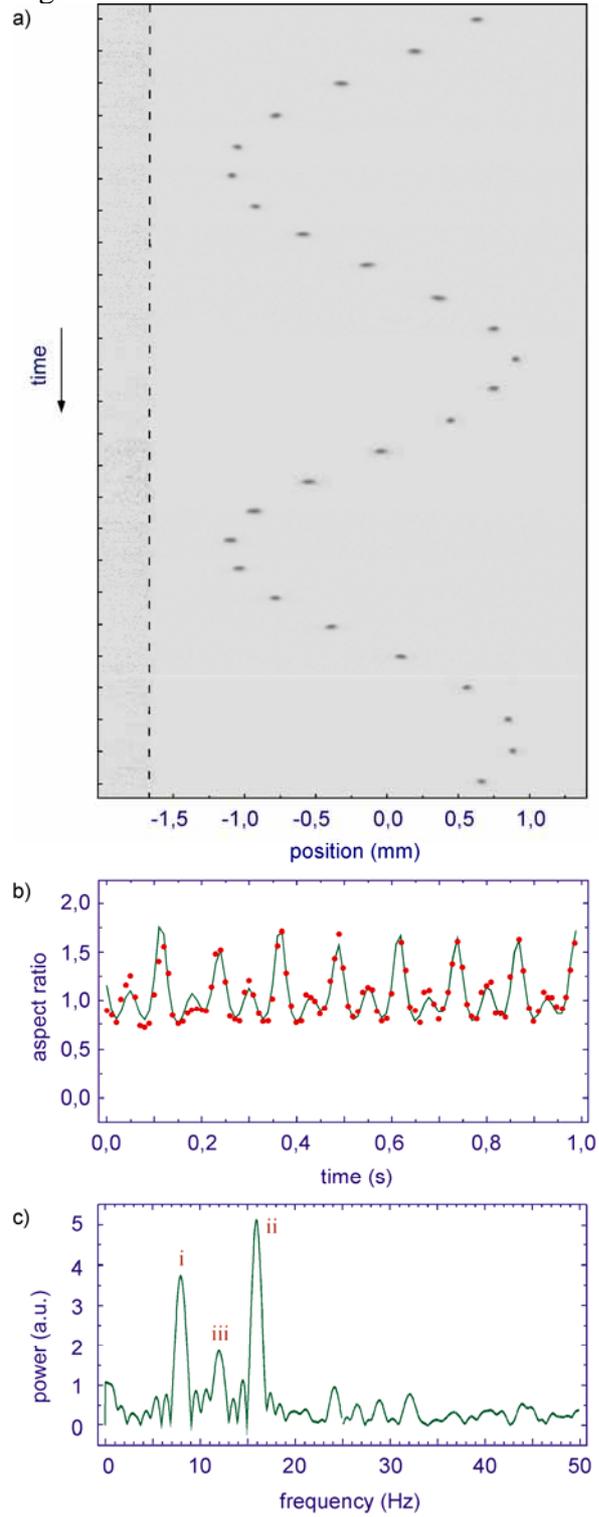

Figure 8

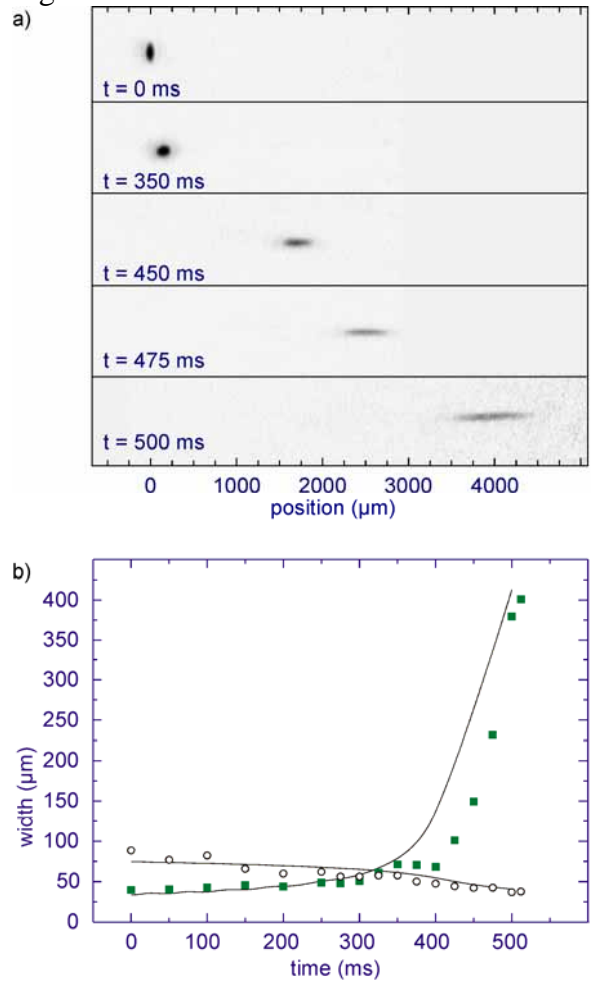

Figure 9

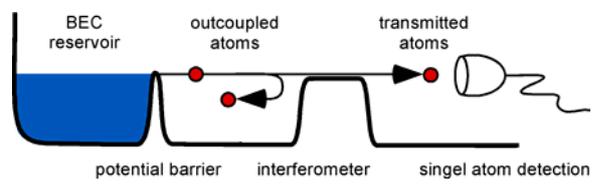